# Growth, Structure and Properties of BiFeO$_3$-BiCrO$_3$ Films obtained by Dual Cross Beam PLD

R. Nechache[1], C. Harnagea[1], L. Gunawan[3], L.-P. Carignan[2], C. Maunders[3], D. Ménard[2], G. A. Botton[3], and A. Pignolet[1]

[1] INRS - Énergie, Materiaux et Télécommunications, Boulevard Lionel-Boulet,
Varennes (Montreal Metropolitan Area), Québec, J3X 1S2, CANADA
[2] Department of Engineering Physics, École Polytechnique de Montréal,
P.O. Box 6079, Montréal, Québec, H3C 3A7, CANADA
[3] Dept. of Materials Science and Engineering, McMaster University,
Hamilton, Ontario, L8S 4M1, CANADA

*Abstract* — The properties of epitaxial Bi$_2$FeCrO$_6$ thin films, recently synthesized by pulsed laser deposition, have partially confirmed the theoretical predictions (i.e. a magnetic moment of 2 $\mu_B$ per formula unit and a polarization of ~80 $\mu$C/cm$^2$ at 0K). The existence of magnetic ordering at room temperature for this material is an unexpected but very promising result that needs to be further investigated. Since magnetism is assumed to arise from the exchange interaction between the Fe and Cr cations, the magnetic behaviour is strongly dependent on both their ordering and the distance between them. We present here the successful synthesis of epitaxial Bi$_2$Fe$_x$Cr$_y$O$_6$ (BFCO x/y) films grown on SrTiO$_3$ substrates using dual crossed beam pulsed laser deposition. The crystal structure of the films has different types of (111)-oriented superstructures depending on the deposition conditions. The multiferroic character of BFCO (x/y) films is proven by the presence of both ferroelectric and magnetic hysteresis *at room temperature*. The oxidation state of Fe and Cr ions in the films is shown to be 3+ only and the difference in macroscopic magnetization with Fe/Cr ratio composition could only be due to ordering of the Cr$^{3+}$ and Fe$^{3+}$ cations therefore to the modification of the exchange interaction between them.

## INTRODUCTION

Multiferroic materials, in which both ferromagnetic and ferroelectric orders coexist, have recently attracted significant attention because of their potential for use in technological applications as well as for fundamental research. Despite their potential usefulness, multiferroic materials are rare in nature and none of them have been used in practical applications so far because few of the known multiferroics has simultaneously a strong response to both external magnetic and electric fields at room temperature.

An example of such a multiferroic material is BiFeO$_3$ (BFO) [1,2]. Bulk BFO intrinsically shows strong ferroelectricity but weak ferromagnetism at room temperature. The single-phase perovskite BiFeO$_3$ is ferroelectric (T$_C$ ~ 1103 K) and antiferromagnetic (AF) (T$_N$ ~ 643 K), exhibiting weak magnetism at room temperature due to a residual moment from a canted spin structure[1]. The bulk single crystal form has been extensively studied [3,4,5,6,7], and it has been shown to possess a rhombohedrally distorted perovskite structure (*a= b= c= 5.63 Å , α= β= γ = 59.4º*) at room temperature. Reports on the properties of BFO in thin film form are quite contradictory, measured values of the spontaneous magnetization ranging from 0 to 0.5 µB per formula unit. [2,8,9,10]. The highest values most probably result from the presence of iron oxides foreign phases formed as byproducts during the deposition process [11] or even by the decomposition of BFO under certain external conditions [12].

Another perovskite, $BiCrO_3$ (BCO) was first synthesized in 1968 using very high pressure firing and was reported to be antiferromagnetic (AF) below 123K with a weak parasitic ferromagnetic moment [13]. A magnetic anomaly was observed around 95K as an increase of magnetic moment with decreasing temperature, which was interpreted as a consequence of a change in the direction of the magnetic easy axis. A first-order structural phase transition was observed around 410K. Below this temperature, the structure is triclinic with $a= c= 3.906$ Å, $b= 3.870$ Å, $\alpha= \gamma= 90.5º$, $\beta = 89.1º$ (at 300K) while above this transition temperature it is pseudomonoclinic with $a= c= 3.878$ Å, $b= 7.765$ Å, and $\beta = 88.8º$ (at 460K). However, the dielectric properties of BCO have not been investigated extensively until recently. Hill *et al.* predicted by first principles calculations that BCO has a G-type antiferromagnetic ground state, with an antiferrodistortive or antiferroelectric structural distortion, similar to that seen in $PbZrO_3$ [14]. Recently, several groups reported experimental results and discussed ferroelectricity in BCO based on anomaly in the temperature dependence of the dielectric permittivity and on reversible switching observed with piezoresponse force microscopy [15,16]. There has been, however, no further conclusive evidence of ferroelectricity in this compound beyond the observation of dielectric anomalies and local piezoelectricity. Meanwhile, D. H. Kim *et al.* [17] have reported that BCO actually exhibits antiferroelectric properties based on well-defined double hysteresis loops observed in the electric field dependence of the dielectric permittivity at 10K and polarization at 15K for epitaxial BCO films capacitors.

Although significant materials developments are still required to generate magnetoelectric materials that could make a real contribution to real appliations such as devices for sensing or data strorage industry, the study of novel multiferroic materials remains important for disruptive technologies ultimately to emerge. In this context, Baettig and Spaldin, using ab-initio first-principles density functional theory, reported the design and calculated properties of a new multiferroic material, namely $Bi_2FeCrO_6$ (BFCO) [18]. They predicted that BFCO would have a rhombohedral structure very similar to that of BFO, except that every second iron cation in the (111) direction is replaced by a chromium cation. The unit cell of BFCO would therefore be a double-perovskite, containing one BFO pseudo-cubic unit cell and one BCO pseudo-cubic unit cell sharing one (A-site) bismuth ion. They also predicted that this new multiferroic material, BFCO, would be ferrimagnetic at 0 K with a magnetic moment of 2 $\mu_B$ per formula unit and would be ferroelectric with a polarization of 80 µC cm$^{-2}$.

Shortly after the prediction by first principles calculation, BFCO epitaxial films grown by conventional Pulsed Laser Deposition (PLD) have been fabricated and were reported to be multiferroic even at room temperature [19]. The films grown on epitaxial $SrRuO_3$ layers or directly on STO were demonstrated to have the expected chemical composition and a correct cation stoichiometry [20]. Contrary to recent reports on BFCO ceramics and thin films [21,22], a clear cation ordering in BFCO epitaxial thin films has recently been demonstrated, i.e. they indeed posses a double-perovskite unit cell with (111) Fe planes alternating with (111) Cr planes [23]. Since magnetism is assumed to arise from the exchange interaction between the Fe and Cr cations, the BFCO films with cationic ordering exhibit a significant magnetization [20,23], in contrast to BFCO without Fe/Cr ordering, which show no magnetic properties [21,22]. Therefore, an important parameter governing the magnetic behaviour is the distance between the $Fe^{3+}$ and $Cr^{3+}$ cations.

Here we attempt to investigate the effect of this parameter by modifying the succession of Fe/Cr cations, i.e. by inducing a controlled "disorder". In order to do this, we grew films with different Fe/Cr cation ratio by simultaneously depositing BFO and BCO by PLD from two different BFO and BCO targets at different deposition rates, using a distinct approach, namely the PLD co-deposition technique dubbed double-crossed-beam PLD (DCB-PLD).

## EXPERIMENTAL

DCB-PLD is an unconventional laser ablation technique previously used to deposit 2-component alloys with controlled composition while avoiding droplet formation [24]. We optimized the experimental setup for low oxygen ambient pressure, conditions that are suitable to have high deposition rates in order to limit the desorption from the surface of metallic Bi, a compound with a high vapour pressure. The geometry of the experimental setup is shown in Fig. 1. The iron and chromium film content is controlled by the fluence of two synchronized laser beams (KrF excimer pulsed lasers having a wavelength of 248 nm) focused on two Bi-rich ceramic targets of $Bi_{1.1}FeO_3$ and $Bi_{1.1}CrO_3$. The bismuth content in the film is mainly controlled by the oxygen pressure in the chamber, while the substrate temperature governs the surface mobility of the incoming species, and hence promotes the epitaxial growth of the film. For comparison, BFCO (equivalent to BFCO 1/1) films have also been grown both by DCB-PLD and by conventional PLD from a single stoichiometric $Bi_2FeCrO_6$ ceramic target as described in earlier reports [19,20,23]. All films were deposited on commercially available (100)- and (111)-oriented Nb-doped $SrTiO_3$ substrates (hereafter STO(100) and STO(111)) to promote the epitaxial growth and control the orientation of the films.

The films crystal structure was investigated by X-ray diffractometry (XRD) θ-2θ scans, Phi-scans as well as reciprocal space mapping (RSM). HRTEM images were obtained at the Canadian Centre for Electron Microscopy with a recently installed and newly developed FEI Titan 80-300HB microscope equipped with a CEOS aberration corrector providing 0.75 Angstrom information limit (FEI, Hillsboro, OR). Their composition was studied by Rutherford Backscattering Spectrometry (RBS) ($He^{++}$ incident beam at 3.068 MeV) and by energy dispersive X-ray spectroscopy (EDXS) in a scanning electron microscope (SEM). Film densities were calculated using X-Ray Reflectometry (XRR). The surface morphology of the films was investigated using scanning electron microscopy (SEM) and atomic force microscopy (AFM). In addition to the structural properties of the PLD-grown epitaxial layers realized, their functional properties were thoroughly investigated: The local ferroelectric properties of the BFCO films were studied by piezoresponse force microscopy (PFM) [25,26,27] that was used to image and manipulate the ferroelectric polarization of the obtained films. Here we used a DI-Enviroscope AFM (Veeco Instruments, Woodbury, NY) equipped with a NSC36a (Micromasch, Tallinn, Estonia) cantilever and tips coated with Co/Cr. We applied an ac voltage of 0.5V at 26 kHz between the conductive tip and the STO:Nb conducting substrate located beneath the BFCO layer and we detected surface induced piezoelectric vibrations using a Lock-in Amplifier from Signal Recovery (model 7265, Wokingham, UK). Field dependent magnetic measurements were obtained for the various compositions, using a vibrating sample magnetometer (VSM). A maximum saturating magnetic field of 10 kOe was applied parallel to film plane, and then decreased down to -10 kOe by step of 500 Oe, and back up to 10 kOe. The field steps were decreased down to 50 Oe around the zero field range for better resolution. An average factor of 20 measurements per field points provided an absolute sensitivity of about $10^{-6}$ emu. The magnetic responses of the sample holding rods and of bare substrates (without BFCO) were also measured and subtracted in order to extract the magnetic signal was originating from the BFCO films.

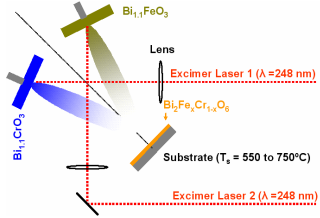

***Figure 1:*** *Schematic of the dual crossed beam pulsed laser deposition geometry (DCB-PLD).*

**RESULTS**

Fig. 2 shows RBS spectra measured for the films with average compositions BFCO 1/3 (i.e. $Bi_2Fe_{1/2}Cr_{3/2}O_6$ or $Bi_4FeCr_3O_{12}$) and BFCO 1/1 (or $Bi_2FeCrO_6$). Using the densities calculated from XRR patterns (8 g/cc for BFCO 1/3 and 8.5 g/cc for BFCO 1/1) we estimate the thicknesses from the films to be about ~ 200 nm and ~ 250 nm respectively. X-ray diffraction θ-2θ spectra show that the films are highly (100)- and (111)- oriented.

When BFCO epitaxial films are grown from a stoichiometric single target by conventional PLD on STO(100) and on STO(111), the distinctive crystal structure of the film consisting in a double perovskite unit cell with B- site $Fe^{3+}/Cr^{3+}$ cation ordering along the [111] direction gives rise to super-lattice peaks in the XRD spectra [23]. Remarkably, we obtain a similar super-lattice pattern for the films BFCO 1/1, grown by DCB-PLD on STO(111), suggesting that the crystal structure of the two BFCO phases (conventional BFCO obtained by conventional PLD from a single BFCO target and BFCO 1/1 obtained by DCB-PLD from two $Bi_{1.1}FeO_3$ and $Bi_{1.1}CrO_3$ targets) is similar, and have indeed a double-perovskite unit cell with Fe/Cr cationic ordering along [111]. We can thus infer that the $Bi_2FeCrO_6$ phase is stable and that its formation is favoured once the optimal growth conditions are found. The different intensities of the superlattice peaks in the case of conventional PLD and DCB-PLD is presently under investigation and might arise from a different "off-the-111-axis" position of the Fe and Cr cations in both cases or from ordered chemical defects (such as oxygen or bismuth vacancies) [28].

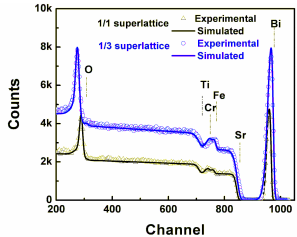

***Figure 2:*** *RBS spectra of BFCO 1/3 and BFCO 1/1 compounds.*

Similarly, super-lattice reflexions are seen in the XRD spectra of the BFCO 1/3 grown by DCB-PLD. In order to study these superlattice reflexions in more detail, we recorded reciprocal space maps near the 111 reflexions of BFCO 1/3 and STO as shown in Fig. 3b. The RSM confirm the presence of the additional superstructure satellite peaks (labeled -2,-1 and +1,+2, in Fig. 3c) in addition to the BFCO 1/3 (111) / STO(111) reciprocal lattice points (RLP) ). The angular separation between the satellite peaks is given by [29]:

$$2\Lambda (\sin \theta_n - \sin \theta_0) = \pm n\lambda, \qquad (1)$$

where $\Lambda$ is the period of the superlattice (SL), $\theta_n$ is the diffraction angle for the $n^{th}$ order of diffraction, $\theta_0$ is the angle for the zero$^{th}$ order ( n = 0) peak, and λ is the Cu K$_\alpha$ X-ray wavelength of 1.5406 Å. Using the formula (1) , we calculate the periodicity of the superlattices and we find them to be ~ 9.3 Å for BFCO 1/3 and ~ 4.6 Å for BFCO 1/1 films.

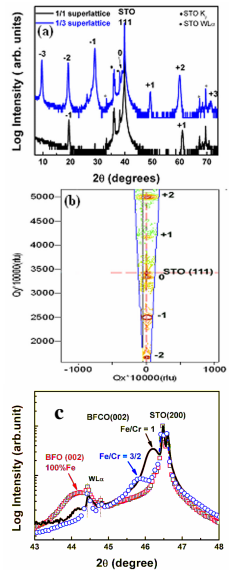

***Figure 3**: (a) X- ray diffraction patterns of BFCO x/y grown directly on (111)-oriented STO:Nb. (b) Reciprocal space mapping (RSM) around the (111) STO reflexion. (c) Sections of X-ray diffraction spectra of BFCO x/y deposited on (100)-oriented STO:Nb, showing the variation of the pseudo-cubic out-of-plane lattice parameter for various Fe/Cr ratios.*

We observe that these superlattice periods correspond to respectively four and two times the pseudo-cubic (111) inter-planar distance $d_{111}$ = 2.3 Å, and reveals the existence of a 1:3, respectively 1:1 Fe/Cr B-site cationic ordering along the [111] crystallographic direction. This is a remarkable result and clearly demonstrates the versatility of the DCB-PLD technique in preparing double perovskite materials with various composition and cation ordering along the [111] direction.

We also studied the effect of the Fe/Cr ratio on the lattice constant of the film. This effect can be estimated from Fig. 3c, showing a limited section of the θ-2θ XRD spectrum of BFCO x/y films grown on (100)-oriented STO:Nb around their 002 reflexions. The variation of the pseudo-cubic lattice parameter is illustrated for three Fe/Cr ratios, namely x/y = 1/0 (i.e. $BiFeO_3$), 3/2, and 1/1 (i.e, $Bi_2FeCrO_6$). As it can be seen in Fig. 3c, the out-of-plane lattice parameter of the BFCO x/y films decreases with decreasing the x/y = Fe/Cr ratio, in agreement with the well-known fact that the bulk lattice parameter of $BiCrO_3$ (3.90 Å) is smaller than that of $BiFeO_3$ (3.96 Å). Additionally, Phi-scans XRD (not shown) performed for these specific films demonstrate that, in the three cases, the film grows epitaxially grown on the substrate.

Figure 4 (a) shows a cross-sectional transmission electron microscopy image of an epitaxial BFCO (1/3) on STO (111) film observed along the [1-10] direction. The epitaxial growth is clearly visible, with the (111) planes parallel to the surface and columnar epitaxial grains extending through the whole film thickness. Small angle grain boundaries are visible between the grains that are not disturbing the epitaxy. Images present some evidence of ordering but further work is required to interpret quantitatively the HRTEM images and the ordering observed. The selected area electron diffraction pattern (SAED) (not shown here), obtained from the cross section confirms the good single crystalline quality of the BFCO layer. The area delimited by the square in Fig 4a is magnified in Fig. 4b, clearly showing a superlattice with a periodicity of 9.4 Å.

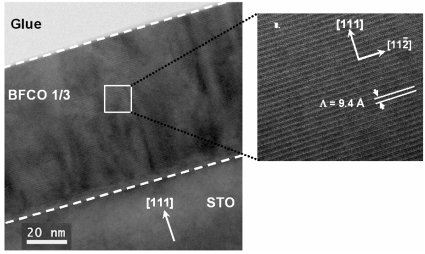

*Figure 4* (a) Cross-section TEM image showing an epitaxial BFCO (1/3) layer grown by DCB-PLD on STO (111) substrate. The direction of observation is [1-10]. (b) Zoom of the area marked by a square in (a) showing a 9.4 Å periodicity.

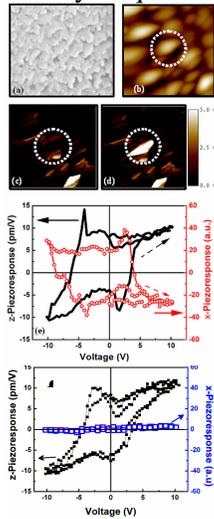

*Figure 5:* (a) SEM image of the surface of a BFCO 1/3 film obtained by DCB-PLD (b-e) Local ferroelectric characterization of a BFCO 1/3 thin film. (b) AFM topography of the film surface. z-PFM images (c) before and (d) after the switching of the grain encircled. Scan size of the images is 2.5×2.5$\mu m^2$. (e) Corresponding in field hysteresis loops of the z-PFM (-) and x-PFM (-o-) signals. (f) In-field PFM hysteresis loops of a (111)-oriented BFCO 1/1 film.

The SEM image in Fig. 5a shows the topography of the surface of one BFCO 1/3 film obtained by DCB-PLD, exhibiting terraces which suggest a layer by layer mode growth consistent with the finding from XRD analysis that (111)-oriented films are epitaxial.

Ferroelectricity of a BFCO 1/3 film grown on a STO(111) substrate at the grain level is demonstrated by local electromechanical measurements using PFM, as shown in Fig. 5 (b-c-d). The grain encircled showed an initial polarization oriented downward (top to bottom, black in Fig. 5c). After a first switching of polarization obtained by applying a positive voltage pulse (+5V) with the AFM tip fixed in contact with the grain, the area was imaged again. The contrast of the grain has changed, indicating that the polarization has switched and is now oriented upward (Fig. 5d). It also reveals that the reversed polarization is stable at least at the experiment time scale (several hours). Furthermore, the recorded PFM hysteresis loops shown in Fig. 5e exhibit a strong PFM signal comparable to that of PZT and BFO films [30]. We were also able to detect a strong in-plane PFM signal. Since spontaneous polarization is expected to lie along the <111> family of crystallographic directions, this is a quite unexpected result. As shown in Fig. 4e, both the z- and x-PFM hysteresis loops suggest switching of polarization by 180º since both components change signs when the bias voltage is reversed. However, the mere presence of an in-plane component of polarization switching means that the polarization vector is not parallel to the (out-of-plane) electric field applied by the AFM tip, although the out-of-plane direction belongs to the <111> family. The same experiment was performed on a BFCO 1/1 film, also (111)-oriented. Similarly, we detected a strong PFM signal and we were able to switch individual grains. Typical hysteresis loops obtained from such a film are shown in Fig 5f. While we still observe a hysteresis of the in-plane PFM signal, its strength is much lower (almost 7 times) than the strength of the in-plane signal in the BFCO 1/3 film. This suggests that the polarization direction in the BFCO 1/3 film deviates significantly from the [111] crystallographic direction, in contrast to the BFCO 1/1 films, for which this deviation is only minor. Further studies are under way to clarify this observation.

Since BFCO films have been reported to have magnetic properties at room temperature, field dependent magnetic measurements were performed for various compositions, using a vibrating sample magnetometer (VSM).

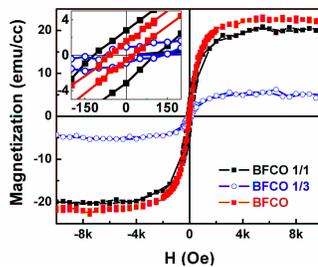

*Figure 6. Magnetic responses from BFCO films grown on 001 Nb doped STO by dual-cross beam PLD with two different superstructures. For comparison, the hysteresis loop obtained for a conventional BFCO is also shown.*

As shown in the inset of Figure 6, epitaxial films obtained by DCB-PLD and having the two types of ordering (BFCO 1/3 and BFCI 1/1) have a clear magnetic hysteresis at room temperature, indicating the presence of an ordered magnetic phase. The saturation magnetization ($M_S$) of the 250 nm thick BFCO (1/1) film was found to be ~ 20 emu/cc, very close to the value obtained for a 300 nm thick BFCO grown by conventional PLD from a single stoichiometric $Bi_2FeCrO_6$ ceramic target. In contrast, a significantly lower saturation magnetization of 5 emu/cc was observed in the 200 nm thick BFCO 1/3 film. Since X-ray photoelectron spectroscopy results indicate that each cation, Bi, Fe and Cr composing the BFCO films were trivalent without revealing the presence of any divalent $Fe^{2+}$ ferrous ions (not shown), a divalent-based mechanism to explain the difference in magnetic properties is ruled out. Therefore, the change in macroscopic magnetization with Fe/Cr ratio composition appears to be due to the ordering of the Cr and Fe cations and therefore to the modification of the exchange interaction between them, in combination with the contribution of the Cr-Cr antiferromagnetic interactions [31].

## CONCLUSION

In summary, $Bi_2Fe_xCr_yO_6$ epitaxial films have been grown on (100)- and (111)-oriented $SrTiO_3$ substrates using a versatile technique which allows the deposition of multicomponent materials with an easy control of the composition, namely dual crossed beam pulsed laser deposition (DCB-PLD). Films of BFCO 1/1 and BFCO 1/3, corresponding to Fe/Cr ratios of 1, respectively 0.33 have been obtained by varying the laser fluence of the two synchronized laser beams. The corresponding XRD θ-2θ spectra exhibit superlattices reflexions that reveal an ordering along the [111] crystallographic direction very similar to the B-site cation ordering recently reported in BFCO [18]. The BFCO 1/1 films obtained by DCB-PLD have a similar composition and XRD signature than the epitaxial BFCO films obtained by conventional PLD. The self-organized growth of BFCO x/y obtained by DCB-PLD in ordered periodic superstructures is presently further investigated in order to better understand, and therefore netter control the structure and the functional properties of the realized multiferroic films.

ACKNOWLEDGEMNTS

The authors want to thank Dr. Martin Chicoine for insightful comments about the RBS analysis. Part of this research was supported by NSERC (Canada), and FQRNT (Québec).